\documentclass[12pt, preprint]{aastex}

\begin{document}

\title{Spitzer Reveals Hidden Quasar Nuclei in Some Powerful FR II Radio Galaxies}

\author{Patrick Ogle}

\affil{Spitzer Science Center, California Institute of Technology, 
       Mail Code 220-6, Pasadena, CA 91125}
\email{ogle@ipac.caltech.edu}

\author{David Whysong\altaffilmark{1} \& Robert Antonucci}

\affil{Physics Dept., University of California, Santa Barbara, CA 93106}

\altaffiltext{1}{now at NRAO, Array Operations Center, P. O. Box O, 1003 Lopezville Rd., Socorro, NM 87801-0387}

\shorttitle{Hidden Quasar Nuclei}
\shortauthors{Ogle et al.}

\begin{abstract}

We present a {\it Spitzer} mid-infrared survey of 42 Fanaroff-Riley class II radio galaxies 
and quasars from the 3CRR catalog at redshift $z<1$. All of the quasars and $45\pm 12\%$ of the narrow-line radio 
galaxies have a mid-IR luminosity of $\nu L_\nu(15 \mu\mathrm{m}) > 8 \times 10^{43}$ erg s$^{-1}$, indicating
strong thermal emission from hot dust in the active galactic nucleus. Our results demonstrate 
the power of {\it Spitzer} to unveil dust-obscured quasars. The ratio of {\it mid-IR luminous} 
narrow-line radio galaxies to quasars indicates a mean dust covering fraction of $0.56 \pm 0.15 $, 
assuming relatively isotropic emission. We analyze {\it Spitzer} spectra of the 14 mid-IR luminous 
narrow-line radio galaxies thought to host hidden quasar nuclei. Dust temperatures of 210-660 K are estimated from 
single-temperature blackbody fits to the low and high-frequency ends of the mid-IR bump. Most of the mid-IR 
luminous radio galaxies have a 9.7 $\mu$m silicate absorption trough with optical depth $<0.2$, attributed to dust 
in a molecular torus. Forbidden emission lines from high-ionization oxygen, neon, and sulfur indicate a source of 
far-UV photons in the hidden nucleus. However, we find that the other $55\pm 13\%$ of narrow-line FR II radio 
galaxies are weak at 15 $\mu$m, contrary to single-population unification schemes. Most of these galaxies are 
also weak at 30 $\mu$m.  Mid-IR weak radio galaxies may constitute a separate population of nonthermal, 
jet-dominated sources with low accretion power. 

\end{abstract}


\section{Unification of Quasars and Radio Galaxies}

The nature of the energy source in active galactic nuclei (AGNs) is a 
fundamental problem. The basic model attributes the large luminosity of these
systems to gravitational energy release in an accretion disk around a 
supermassive black hole. A jet may be driven by magnetic fields threading 
the disk \citep{bp82}. The black hole spin energy may also be tapped and 
converted into electromagnetic Poynting flux and particles in a relativistic 
jet \citep{bz77,pc90,m99,d04}.

Extragalactic radio sources are categorized by their morphology as either of two 
types \citep{fr74}. Fanaroff-Riley (FR) type I sources are edge-darkened, while FR IIs
are edge-brightened. The different morphology of FR Is indicates that they are not 
related to FR IIs by orientation. FR Is also have lower radio luminosities than FR IIs 
for a given host galaxy luminosity \citep{ol94,b96}  and most have low-ionization nuclear
emission region (LINER) spectra \citep{hl79}. However, not all FR Is can be characterized by low
accretion power  \citep{wa04,cr04}. The present paper focuses on FR IIs, which contain powerful
jets with bright terminal hot spots and lobes. Furthermore, we count broad-line radio galaxies 
(BLRGs) as low-luminosity quasars.

Quasars and narrow-line radio galaxies (NLRGs) may be unified by orientation-dependent obscuration.
Radio galaxies are thought to host quasar nuclei that are obscured by circumnuclear dusty tori 
aligned with the radio jets \citep{ant84} . Unification of radio galaxies and quasars can therefore 
explain the lack of quasars viewed at large angles to the radio axis \citep{b89}. The percentage 
of high-redshift radio galaxies (60\% of the 3CRR FR II sample at $z>0.5$) would then indicate a torus 
covering fraction of $\sim 0.6$.

However, there appears to be a discrepancy between the redshift distributions of quasars and
radio galaxies at $z<0.5$, with a factor of $\sim 4$  more narrow-line radio galaxies 
than quasars \citep{s93}. Furthermore, the median projected linear size of
these 'excess' radio galaxies is smaller than expected for quasars seen 
in the sky plane \citep{s93,w05}. The unification hypothesis may be
modified to include a second population of lower luminosity, low-excitation
FR II radio galaxies \citep{wj97,grw04}. Alternatively, it has been argued that the 
torus covering fraction may increase with decreasing radio luminosity \citep{l91}.

The unification hypothesis has been qualitatively confirmed by 
spectropolarimetry of radio galaxies, many of which have been shown
to have highly polarized broad emission lines and blue continuum, scattered from material 
which has a direct view of the active galactic nucleus \citep{cdb97, cot99}. Of particular note
are the original discovery of highly polarized broad H$\alpha$ from the hidden quasar nucleus in 
3C 234 \citep{ant84}, and the discovery of highly polarized broad H$\alpha$ in the 
spectrum of the powerful radio galaxy Cygnus A \citep{ocm97}. However, this method of detecting 
hidden quasars relies on an appropriately placed scattering region to view the otherwise 
hidden nucleus. Such a region is not guaranteed to exist for all radio 
galaxies, and thus spectropolarimetry can easily yield false negatives. 
Polarimetry is also ineffective at determining the luminosity of the hidden 
nucleus, since the scattering efficiency is usually unknown.

Another way to search for hidden quasar nuclei is to observe radio 
galaxies in the mid-IR. If the unification hypothesis is correct, the dusty torus should serve 
as a crude calorimeter of the central engine \citep{mhm01,sfk04,hmb04,wa04}. 
Optical, UV, and X-ray  photons from the quasar nucleus are absorbed by dust in the 
torus and the energy is re-emitted in the thermal infrared. This explains why blue, UV color-selected 
quasars emit 10-50\% of their luminosity in the IR \citep{spn89,hmc00}. There appears 
to be no connection between the bulk of this IR emission and nonthermal radio emission, 
except in core-dominated radio sources such as blazars. Observations of matched 3CR quasars and 
radio galaxies by ISO indicate similar IR luminosities, consistent with the unification
picture \citep{mhm01,hmb04}. However, differences in 24 $\mu$m/ 70 $\mu$m color
may indicate that mid-IR emission from the torus is anisotropic by a factor of $\le 3$
\citep{srh05}. 

We present {\it Spitzer} observations of a sample of 42 FR II radio galaxies and quasars selected 
from the 3CRR survey. The goals are to search for mid-IR emission from hidden quasar nuclei and test 
the ubiquity of the unification hypothesis. The {\it Spitzer} Infrared Spectrograph (IRS) combines 
the advantages of unprecedented sensitivity from 5-36.5 $\mu$m to measure the mid-IR continuum and 
spectral resolution to measure high ionization emission lines powered by hidden AGNs. 
In the current paper, we present evidence for hidden quasar nuclei based on mid-IR photometry extracted from 
the IRS spectra. We examine in detail the spectra of the subset of 14 mid-IR luminous radio galaxies which
appear to contain hidden quasar nuclei. Spectra of the quasars and mid-IR weak radio 
galaxies and a statistical study of the complete sample will be presented in separate papers.

\section{Sample}

We begin by selecting a well-defined, radio flux-limited and redshift-limited sample of 55 
radio galaxies and quasars from the 3CRR catalog \citep{lrl83}. We include all 3CRR sources with 
FR II radio morphology, a flux of  $S_{178}>16.4$ Jy\footnote{The radio flux limit is 15 Jy using
\cite{lrl83} flux values and 16.4 Jy on the standard \cite{b77} scale.} at 178 MHz, and a redshift of
$z<1$. The original 3CRR catalog has a flux limit of 10 Jy at 178 MHz, is restricted to northern 
declinations ($\delta >10 \arcdeg$), and has galactic latitude $|b|>10\arcdeg$. It is the canonical 
low-frequency selected catalog of bright radio sources, has optical identifications and redshifts 
for all entries, and has been extensively observed in most wavebands. 

We select only sources with FR II radio morphology. We verify or update the FR classification of 
all sources by inspection of the latest published radio maps. Compact, steep-spectrum sources 
(CSSs: 3C 48, 138, 147, 286, and 309.1) with radio major axis $D<10$ kpc \citep{ffp85} are excluded 
from the sample because they may constitute a class of young or frustrated radio sources. Here and 
throughout this paper, we assume a cosmology with $H_0=70$ km s$^{-1}$ Mpc$^{-1}$, 
$\Omega_\mathrm{m}=0.3$, and $\Omega_\Lambda=0.7$. Size and morphology indicate that CSSs are not 
related to FR IIs by orientation. 

It is essential for our unification studies that we select a sample based on isotropic
radio lobe flux, and {\it not} on optical or IR properties, so that it is unbiased by 
orientation-dependent selection effects. In particular, our sample includes no blazars. No sources
make the flux limit only because of beamed emission from the core of the radio jet. Our sample 
includes quasars as well as radio galaxies, and we use the quasar subsample as a control. We aim to
determine whether and which narrow-line FR II radio galaxies have mid-IR power comparable to 
quasars or broad-line radio galaxies of similar radio lobe flux and redshift.

The 42/55 sources in our sample which we have observed with {\it Spitzer}, or which have {\it Spitzer} data in the
public archive are listed in Tables 1 and 2. The 25 {\it mid-IR luminous} sources with 
$\nu L_\nu(15\mu\mathrm{m}) >8\times 10^{43}$ erg s$^{-1}$ (14 NLRGs and 11 quasars or BLRGs) are listed 
in Table 1, and the 17 {\it mid-IR weak} galaxies with $\nu L_\nu(15\mu\mathrm{m}) <8\times 10^{43}$ erg s$^{-1}$ 
are listed in Table 2. The reason for this particular division is explained below. 

Optical source classifications are based on emission line properties.
Type 1 sources have directly visible broad emission lines (quasars and BLRGs), and type 2 sources
(NLRGs) do not. The NLRGs are further classified using their forbidden emission lines 
\citep{jr97,wrb99}\footnote{Updated optical classifications are available at 
http://www-astro.physics.ox.ac.uk/$\sim$cjw/3crr/3crr.html.}.  
High-excitation galaxies (HEGs) are defined  to have [O {\sc iii}] $\lambda$5007 
equivalent widths of $>10$ \AA~ and [O {\sc iii}] $\lambda$5007/[O {\sc ii}] $\lambda$3727 $>1$. The 
sources which do not meet these criteria are classified as low-excitation galaxies (LEGs).
The equivalent width criterion ensures that [O {\sc iii}] is measurable in moderate S/N spectra. However, 
it remains to be seen whether some sources with low [O {\sc iii}] equivalent width might have 
[O {\sc iii}] $\lambda$5007/[O {\sc ii}] $\lambda$3727 $>1$. In addition, we caution that [O {\sc ii}] 
and [O {\sc iii}] may be subject to differing amounts of extinction.

\section{Observations}

We observed the sources in our sample with the Infrared Spectrograph (IRS) on the 
{\it Spitzer} Space Telescope \citep{h04,w04}. We used the low-resolution ($R\sim 64-128$) modules 
Short-Low (SL) and Long-Low (LL) for accurate spectrophotometry over the wavelength
range 5-36.5 $\mu$m. Wavelengths 36.5-40 $\mu$m are unusable because of 
low-S/N and 2nd-order bleed-through caused by filter delamination in LL 1st order (LL1). The 
absolute and relative flux accuracies of IRS are generally better than 10\% and 4\%, respectively, as 
judged from observations of bright standard stars. However, additional low-level
instrumental artifacts may become important for faint sources.

We used IRS in standard flux-staring mode, for 2 cycles at 2 nod positions in each 
of the modules SL1, SL2 (SL 1st and 2nd order), and LL2 (LL 2nd order). We executed
1 cycle at 2 nod positions for LL1, which covers the 20-36.5 $\mu$m range. A typical 
observation includes 240 s of on-source exposure time in each of SL1, SL2, and LL1, and 
480 s in LL2, for a total of 2000 s per target (including overhead).
Archived IRS data are used for 14 sources which were observed partly or in full by other 
investigators (with similar or longer exposure times). 

Nod or off-slit observations were subtracted to remove foreground emission from the 
telescope, zodiacal light, and interstellar medium. Spectra were then extracted from the 
Basic-Calibrated Datasets (BCDs), using the {\it Spitzer} IRS Custom Extraction 
(SPICE\footnote{ http://ssc.spitzer.caltech.edu/postbcd/spice.html} version 1.1) software 
and standard tapered extraction windows. The extraction window full-widths are proportional to 
wavelength in each order to match the diffraction-limited telescope point-spread function 
(SL2: $7\farcs2$ at 6 $\mu$m, SL1: $14\farcs 4$ at 12 $\mu$m, LL2: $21\farcs7$ at 16 $\mu$m,
LL1: $36\farcs6$ at 27 $\mu$m). We rebinned portions of the spectra of 3C 55, 172, 220.1, 244.1, 263.1, 
280, and 330 by factors of 4-8 in order to improve the S/N at short wavelengths. Spectral 
orders were trimmed at the edges and merged to produce final spectra. 

The SL and LL slits have widths of $3\farcs7$ and $10\farcs6$, respectively. Standard point-source flux 
calibrations (version 12.0) were applied to correct for slit and aperture losses and convert the spectra from 
electron s$^{-1}$ to Jy. In most cases, fluxes match to $<15\%$ across order boundaries, consistent with a 
point source that is well-centered in all of the slits. However, in 5 cases (3C 192, 216, 220.1, 380, and 
381) SL2 fluxes are larger by 17-35\% relative to the other orders. Assuming that these mismatches owe 
to variable slit-loss caused by pointing errors,  the orders with low flux are adjusted upward to match
the orders with high flux. Order mismatches may alternatively be an indication of extended mid-IR 
emission. 

The results for a few sources with nearby neighbors in the slit should be viewed with caution.
In the case of 3C 310, a nearby companion galaxy (to the east) may contribute a significant fraction of the
flux ($<50\%$) in the LL1 slit. Similarly, a nearby source may potentially contribute to the LL spectra 
of 3C 438 (which is, however undetected at 15 $\mu$m). The SL2 spectrum of 3C 388 may be weakly affected by flux 
from a nearby star on the slit ($<20\%$). The northern component of the double nucleus in 3C 401 falls 
outside of the SL slits, but falls inside the LL slit used to measure the 15 $\mu$m flux.

\subsection{Mid-Infrared and Radio Luminosities}

We measure the mean $6.5-7.5$ $\mu$m and $13.0-17.0$ $\mu$m flux densities $F_\nu(7$ and 15 $\mu$m, 
rest) of each target (Tables 1 \& 2). All {\it Spitzer} flux densities in this paper are in observed units at
a constant rest-frame wavelength defined by $\lambda_\mathrm{rest}=\lambda_\mathrm{obs}/(1+z)$, where 
$z$ is the redshift measured from optical emission lines and cataloged in the NASA Extragalactic 
Database (NED\footnote{ http://nedwww.ipac.caltech.edu}). This avoids any complication from potentially
large cosmological K-corrections that could otherwise be introduced by a steep IR continuum slope or 
redshifted silicate absorption features. 

We choose to measure the mid-IR flux at 7 and 15 $\mu$m to avoid the 9.7 $\mu$m trough and the deepest part of the 
18 $\mu$m silicate absorption trough. We exclude the 14.0-14.5 $\mu$m and 15.3-15.8 $\mu$m wavelength regions from 
our photometry, to avoid emission from Ne {\sc v} and Ne {\sc iii}.  The 7 and 15 $\mu$m bands are within the 
Spitzer IRS bandpass for redshifts $z<1.28$. However, the archival LL data for two quasars (3C 254 and 275.1) are 
not yet public. We extrapolate their SL spectra to obtain $F_\nu($15 $\mu$m, rest) using $F_\nu($7 $\mu$m, rest) and 
the observed (relatively line-free) 5-7 $\mu$m spectral index. 

Radio luminosities $\nu L_\nu($178 MHz, rest) are estimated from the observed 178 MHz fluxes  
and K-corrected using the 178-750 MHz radio spectral index \citep{lrl83}. The sources in our sample
display a large range of nearly 3 orders of magnitude in mid-IR to radio luminosity: 
$\nu L_\nu(15$ $\mu\mathrm{m})/\nu L_\nu(178$ MHz$)=0.8-680$ (Fig.1). This quantity is thought to reflect the
relative importance of accretion luminosity and jet kinetic power dissipation. However, different
size and time scales are probed by the radio (10 kpc-1 Mpc) and mid-IR (0.1-100 pc), and the radio
power may be sensitive to differences in environmental conditions.

For the purpose of studying quasar and radio galaxy unification, it is natural to divide the sample into
the {\it mid-IR luminous} NLRGs which emit as powerfully as quasars or BLRGs, and the {\it mid-IR weak} 
NLRGs that do not. We adopt an empirical dividing line of $\nu L_\nu(15$ $\mu\mathrm{m})> 8 \times 10^{43}$ 
erg s$^{-1}$ to separate hidden quasars from mid-IR weak radio galaxies. The cutoff is set at 1/2 the 
luminosity of the mid-IR weakest BLRG (3C 219) to allow for some degree of anisotropy at 15 $\mu$m.
Fourteen NLRGs satisfy our criterion and are thus likely to contain hidden quasar or BLRG 
nuclei (Table 1). Notably, all of these NLRGs are optically classified as HEGs.

The 17 mid-IR weak NLRGs with $\nu L_\nu(15$ $\mu\mathrm{m})<8 \times 10^{43}$ erg s$^{-1}$ (Table 2) 
have mixed optical classifications, including both HEGs and LEGs. These sources have lower S/N mid-IR 
spectra, which will be considered in detail in a later paper. Six mid-IR weak NLRGs (including 2 HEGs 
and 4 LEGs) are undetected by {\it Spitzer} at 15 $\mu$m, and one is also undetected at 7 $\mu$m.

\subsection{Hidden Quasar Spectra}

\subsubsection{Continuum Emission}

We now present {\it Spitzer} spectra of the 14 mid-IR luminous NLRGs that ostensibly contain
hidden quasar nuclei (Figs. 2-4). We also plot the spectral energy distributions (SEDs) of the sources
with published near-IR photometry (Fig. 5). The collected photometric data were measured in the J, H, 
K, L$^\prime$, and M wavelength bands from the ground \citep{ll84,llm85,srl99,sww00}.
The photometric apertures range in size from $3-11\arcsec$, with preference given to the apertures
that most closely match the {\it Spitzer} SL slit width. Where available, the ground-based L$^\prime$ 
and M-band photometry agrees with {\it Spitzer} spectrophotometry remarkably well. There is no 
indication of variability over the time span of 20 yr.

Four of the low-redshift NLRGs (3C 33, 234, 381, and 452) have broad peaks in their $\nu L_\nu$ spectra (and 
SEDs) at $1.5-2.5\times 10^{13}$ Hz (12-20 $\mu$m). A maximum and spectral curvature near 20 $\mu$m are 
also suggestive of broad peaks in the {\it Spitzer} spectra of 3C 55, 244.1, 265, and 330.
The large amplitude ($\sim 0.5-1.0$ dex) of the mid-IR bump (Fig. 5) excludes a large contribution of 
synchrotron emission to the mid-IR continuum of most sources. This is not surprising if the equatorial
plane of the dusty torus is roughly perpendicular to the radio jet, such that jet emission is beamed 
away. The high redshifts of the NLRGs 3C 172, 220.1, 263.1, 268.1, and 280 preclude
the identification of a mid-IR bump in the SEDs of these sources. The unusually flat, blue SED of 3C 433 may 
indicate a quasar viewed at {\it low} inclination (Section 3.2.2).

We attribute the mid-IR continuum bump visible in most sources to thermal emission from warm or hot 
dust. Fitting the mid-IR peak with a single-temperature blackbody model indicates  
dust with a temperature of $210-225\pm 0.5$ K (Fig. 5). While this temperature characterizes the peak 
of the mid-IR SED, hotter dust must also be present. At frequencies greater than the peak of the SED 
($2.0-7.5\times 10^{13}$ Hz), the continuum emission of most sources can be characterized using a power law 
with spectral index $\alpha = 1.1-2.1$ (Table 1 \& Fig. 6). This emission likely comes from a continuous 
distribution of dust temperature. We measure the spectral index between 7 and 15 
$\mu$m, avoiding the 9.7 $\mu$m and 18 $\mu$m silicate absorption troughs.  The most blue 
and apparently hottest mid-IR luminous NLRG is 3C 265, while the most red and coolest are 3C 55 and 3C 268.1 
(Fig. 6). In comparison, some mid-IR weak sources such as 3C 310 and 3C 388 are quite blue 
($\alpha \sim -0.1- +0.7$), indicating a large contribution of starlight from the host galaxy to the 
7 $\mu$m continuum.

The near-IR continuum shifts into the {\it Spitzer} IRS passband for the highest redshift ($z>0.7$) sources.
The spectra of the NLRGs 3C 55 and 3C 265 steepen above $7.5\times10^{13}$ Hz (below 4 $\mu$m). Fitting these 
spectral turnovers with single-temperature blackbodies, we find emission from hot dust with temperatures of 
$520\pm 10$ K and $660\pm 10$ K, respectively.  Altogether, the mid-IR luminous radio galaxies in our sample show 
emission from dust with temperatures distributed in the range 210-660 K. Hotter temperature dust (up to the 
sublimation temperature) may be present but not visible for radio galaxy tori viewed at high inclination \citep{pk92}.

Extinction by cold foreground dust in the host galaxy may also affect the spectral index.
For Galactic-type dust, $A(7,15,35~\mu\mathrm{m})/A(\mathrm{V})=(0.020,0.015,0.004)$ \citep{m00}. An 
extinction of $A(\mathrm{V})=100$ would steepen the 7-15 $\mu$m spectral index by $\delta \alpha= 0.6$ 
(Fig. 6a). The observed range in spectral index for the mid-IR luminous NLRGs is $\delta \alpha=1.0$, 
corresponding to $A(\mathrm{V}, 7, 15, 35~\mu\mathrm{m})=(167, 3.3, 2.5, 0.6)$ mag. Thus if reddening 
by a cold foreground dust screen accounted entirely for the range in mid-IR slope, the mid-IR emission 
could be anisotropic by factors of $f_\mathrm{A}(7,15,35~\mu\mathrm{m})\sim (22,10,1.3)$. However, 
these are upper limits since variations in the spectral index are also controlled by the physical 
temperature distribution of the visible dust.

The SEDs of several sources (3C 33, 55, 172, 265, and 452) have upturns at short wavelengths, which we 
attribute to stellar emission from the host galaxy (Fig. 5). The wavelength of the upturn (1-5 $\mu$m) 
is an indicator of the relative strength of the mid-IR bump seen from our direction vs. host galaxy 
light, occurring at shorter wavelength for sources with a stronger mid-IR bump. This may have important 
consequences for understanding the K-z Hubble diagram for 3C radio galaxies, for which it has been 
argued that AGNs contribute a negligible fraction of the K-band flux \citep[e.g.,][]{sww00}. This may 
be incorrect for a few of the most luminous mid-IR sources in our sample, including 3C 234 and 3C 280 
where there appears to be much emission from hot dust in the K band. 

Detailed spectral modeling, combined with radio orientation indicators, promises to further 
characterize the temperature distribution, optical depth, and inclination of the dusty torus that is
thought to produce most of the mid-IR emission from hidden quasar nuclei. Such an analysis is, however,
outside the scope of the present paper.

\subsubsection{Silicate Absorption}

The silicate absorption trough at 9.7 $\mu$m is detected in 12/14 of the mid-IR luminous
NLRG spectra (Table 3 \& Figs. 2-4). The equivalent width EW$_{9.7}$ and apparent optical depth 
$\tau_\mathrm{9.7}$ are measured relative to a local continuum fit to either side of the trough,
indicated in Figures 2-4. The optical depth is averaged over the trough bottom (rest 9.2-10.2 $\mu$m)
to improve the S/N. It should be kept in mind that the apparent $\tau_\mathrm{9.7}$ is just a convenient
parameterization of (and lower limit to) the total optical depth since there must also be broad-band 
silicate absorption of the adjacent continuum.

The apparent silicate optical depths are small ($\tau_\mathrm{9.7}=0.02-0.2$), for all but 3C 55 and 
3C 433. If attributed to foreground dust screens, this would indicate optical extinction of only 
$A_\mathrm{V}=0.2-5.1$  mag (Fig. 6b), assuming a Galactic extinction law with 
$A_\mathrm{V}/\tau_\mathrm{9.7}=12.3-25.6$ mag \citep{rl85,dl84}. The extinction values are clearly 
underestimated since they imply that the hidden nuclei in 3C 234, 265, 381, and 452, which have 
$\tau_\mathrm{9.7}\le 0.1$, should be reddened but directly visible at H$\alpha$. The same discrepancy 
between $\tau_\mathrm{9.7}$ and estimates of extinction at shorter wavelengths is seen for the hidden 
quasar nucleus in Cygnus A, and attributed to a radial gradient in torus dust temperature \citep{iu00}.
The observed range of $\tau_\mathrm{9.7}$ may correspond to a range of equatorial silicate dust column 
densities in the torus, or alternatively a range of viewing angles. In this regard, more detailed 
modeling of the torus, including its geometric and temperature structure is clearly called for.

Filling-in of the silicate troughs by silicate {\it emission} from the torus or 
narrow-line region (NLR) may also reduce the apparent silicate optical depths in some sources. 
This is predicted for an optically thick torus viewed at an intermediate or face-on inclination 
\citep{pk92}. Recently, strong silicate emission features were detected by {\it Spitzer} in several 
radio-loud (3C) and radio quiet (PG) quasars \citep{shk05,hss05}. The failure of previous attempts to 
observe this feature inspired torus models with large dust grain size \citep{ld93} or a spatial 
distribution of optically thick clumps \citep{nie02}. However, it appears that past non-detections owe
to inadequate wavelength coverage to determine the underlying continuum.

The NLRGs 3C 55 and 3C 433 have significantly deeper silicate troughs than other NLRGs, with 
$\tau_\mathrm{9.7}=0.9$ and 0.7, respectively (Fig. 6b). We suggest that their active nuclei and tori are 
absorbed by an additional (kpc-scale) cold dust screen in the host galaxy. As noted above, the NLRG 3C 433 
is unusual in having a flat, blue continuum (similar to some of the quasars in our sample). A blue mid-IR 
spectrum is not necessarily at odds with deep silicate absorption features. It can be understood if this 
is a quasar viewed at low inclination to the jet and torus axes, but through an 
($A_\mathrm{V}\sim 10$) cold dust screen. This amount of extinction would result in very little 
reddening at 7-15 $\mu$m ($\delta \alpha=0.05-0.14$), but would be sufficient to create the deep 9.7 $\mu$m 
trough (Fig. 6b) and would obscure any optical broad lines. 

The NLRG 3C 433 is also unique in having the only unambiguously detected 18 $\mu$m silicate trough, 
with equivalent width $EW_{18}=0.42 \pm 0.01$ $\mu$m and apparent optical depth 
$\tau_\mathrm{18}=0.07 \pm 0.03$ (averaged over 17-19 $\mu$m). The ratio of $\tau_\mathrm{18}$ to 
$\tau_\mathrm{9.7}$ apparent silicate trough depths is $0.10 \pm 0.04$, consistent with a $0.11$ ratio 
for Galactic-type silicate dust \citep{dl84}. We do not see the full 18 $\mu$m silicate trough in the 
spectrum of 3C 55 because of inadequate rest-wavelength coverage.

\subsubsection{Forbidden Emission Lines}

All of the mid-IR luminous NLRGs with high S/N spectra have forbidden emission lines from 
highly ionized metals, including [O {\sc iv}]  $\lambda 25.89$ $\mu$m, [Ne {\sc ii}] $\lambda 12.81$, 
[Ne {\sc iii}] $\lambda 15.55$, [Ne {\sc v}] $\lambda 14.3$, [Ne {\sc v}] $\lambda 24.31$, 
[Ne {\sc vi}] $\lambda 7.65$, [S {\sc iii}] $\lambda 18.71$, [S {\sc iii}] $\lambda 33.48$, and 
[S {\sc iv}] $\lambda 10.51$ (Figs. 2-4). We measure the line flux and rest equivalent 
width of each emission line relative to the local continuum level (Table 4). Formal 
uncertainties are computed from the noise in the continuum to either side of the line. Upper 
limits are estimated for undetected emission lines, assuming they are unresolved. 

The large range of ionization states (especially high-ionization Ne {\sc v}, Ne {\sc vi}, and 
S {\sc iv}) indicates photoionization by a hidden source of far-UV photons \citep{asl99,slv02,acs04}, 
e.g. a quasar nucleus. Low critical densities in the range $10^3-10^6$ cm$^{-3}$ \citep{asl99} indicate 
that the forbidden lines arise in the NLR. There could plausibly be contributions 
from starburst emission to the lower-ionization emission lines such as [Ne {\sc ii}].
For 3C 55 and 3C 433, the [S {\sc iv}] $\lambda 10.51$ line has a relatively large flux even 
though it sits at the bottom of a deep silicate trough. This line must then arise from a region 
not heavily obscured by dust, such as the extended NLR. The resolving power of IRS is 
insufficient to measure the intrinsic emission line widths, which are therefore $<4700$ km s$^{-1}$.
 
In order to assess the ionization level of the emission line regions, we compute several emission 
line ratios (Fig. 7). In particular, the [O {\sc iv}]/[Ne {\sc ii}] and [Ne {\sc v}]/[Ne {\sc ii}] ratios 
can be used as diagnostics of the relative contributions of AGN and starburst emission to the mid-IR emission
line spectra of galaxies \citep{g98,slv02}. In the sources where O {\sc iv} , Ne {\sc v}, and Ne {\sc ii} 
emission are all detected (3C 33, 234, 381, and 433), we find 
[O {\sc iv}]  $\lambda 25.89$ $\mu$m / [Ne {\sc ii}] $\lambda 12.81$ $\mu$m $>1.0$ (Fig. 7a) and 
[Ne {\sc v}] $\lambda 14.3$/ [Ne {\sc ii}] $\lambda 12.81$ $\mu$m $> 0.5$ (Fig. 7c), indicating a $>50\%$ AGN 
contribution to the emission line spectrum. The [Ne {\sc ii}] $\lambda 12.81$ $\mu$m line is relatively weak or 
undetected in the $z>0.2$ sources, making it difficult to apply these diagnostics. However, the large EWs of the 
[Ne {\sc vi}] or [S {\sc iv}] lines in 3C 55, 265, and 330 indicate that the emission line spectra of these sources 
are also AGN-dominated.

\subsubsection{Molecular Emission}

Polycyclic aromatic hydrocarbon (PAH) emission features are commonly seen in star-forming regions, 
starbursts, and starburst-dominated ULIRGs \citep[e.g.,][]{g98,acs04}. The only PAH feature we detect
is the weak 11.3 $\mu$m line in the spectrum of 3C 33, with a flux of $2.3\pm 0.3 \times 10^{-14}$ erg s$^{-1}$ 
cm$^{-2}$ and equivalent width of $0.009\pm 0.003$ $\mu$m (Fig. 2). We do not detect a 6.2 $\mu$m PAH feature in 
3C 33 (EW$<0.8$ $\mu$m) or any of the other mid-IR luminous NLRGs, though this spectral region is generally 
noisier. If present, we could not cleanly resolve the 7.7 $\mu$m PAH feature from the adjacent [Ne {\sc vi}] 
line, nor the 12.7 $\mu$m PAH feature from the adjacent [Ne {\sc ii}] line. Regardless, we do not see any hint 
of these PAH features in any source, suggesting EW$<<0.1$ $\mu$m. Therefore in most cases, neither the 7.7 or 11.3 
$\mu$m PAH features can have a significant impact on the measurement of the silicate trough. 

The general lack of PAH features is a strong indication that the primary power source in mid-IR luminous 
radio galaxies is accretion power, not hot stars. It is likely that PAHs are destroyed in the torus, which is 
exposed to intense X-ray emission from the AGN \citep{v91}. The weak PAH emission that is present in 3C 33 may 
arise in star-forming regions shielded from the AGN.

We detect the H$_2$ 0-0 S(3) 9.67 $\mu$m and H$_2$ 0-0 S(1) 17.03 $\mu$m pure rotational emission lines of 
molecular hydrogen (at the $3\sigma$ level) only in the spectrum of the NLRG 3C 433 (Fig. 2). The line fluxes are 
0.6 $\pm 0.2 \times 10^{-14}$ and 1.4 $\pm 0.4 \times 10^{-14}$ erg s$^{-1}$ cm$^{-2}$ respectively (EW $=$ 0.008 
and 0.013 $\mu$m). The location of the 9.67 $\mu$m line at the bottom the deep 9.7 $\mu$m silicate trough may 
indicate that the H$_2$ emission region is exterior to the obscuring dust screen. The H$_2$ emission lines can be 
produced in warm molecular gas heated either by shocks or X-ray photons from the AGN \citep{rkl02}. The ratio of 
S(3)/S(1) line fluxes is $0.5 \pm 0.2$, which indicates warm H$_2$ with an excitation temperature of $300\pm30$ K. 
We estimate a warm H$_2$ mass of roughly $2\times 10^8 M_\odot$, assuming a Boltzmann distribution of rotational 
level populations, an unresolved source, and negligible mid-IR extinction.

\section{Discussion}

\subsection{Hidden Quasar Nuclei}

The high mid-IR luminosities $\nu L_\nu(15$ $\mu\mathrm{m})= 10^{44}-10^{46}$ erg s$^{-1}$ of 
$45 \pm 12\%$ (14/31) of the FR II NLRGs in our sample are consistent with hidden quasar or BLRG nuclei. In 
fact, such copious hot dust emission directly requires a hidden source of quasar-like luminosity to power it.  
Including the 11 quasars and BLRGs seen directly, we find that at least $60 \pm 12\%$ (25/42) of 3CRR FR II 
sources at $z<1$ with $S_{178}>16.4$ Jy contain powerful AGNs. 

The percentage of {\it mid-IR luminous} AGNs obscured by dust and therefore classified as NLRGs is 
$56\pm 15\%$ (14/25), corresponding to a mean torus covering fraction of 0.56 and mean torus opening 
half-angle of $55\pm 11 \arcdeg$. (If the 8 mid-IR weak HEGs are counted as highly obscured quasars, 
then the mean torus covering fraction increases to $0.67 \pm 0.14$.) Both of these numbers are 
consistent with the estimated 60\% mean torus covering fraction required to unify $z=0.5-1.0$ radio 
galaxies and quasars \citep{b89}.

The receding torus model \citep{l91} predicts a larger torus covering fraction for low luminosity sources.  
If so, we would expect a larger mean mid-IR/radio ratio and a smaller type 1 fraction for low-redshift than for 
high-redshift sources. The fraction of type 1 mid-IR luminous sources is 3/9 ($0.3\pm 0.2$) at $z<0.5$ vs. 8/16 
($0.5 \pm 0.2$) at $z>0.5$. Clearly, a larger sample is needed to put meaningful constraints on the variation 
of torus covering fraction with redshift or luminosity.

For the galaxies that contain a powerful mid-IR source, we find that
there are other indications of a hidden AGN. The high-ionization, mid-IR
forbidden lines such as [Ne {\sc v}], even more-so than the strong optical [O {\sc iii}] lines,
are telltale signatures of a non-stellar source of FUV photons in the radio galaxies that show 
them. Because of the lower extinction in the mid-IR relative to the optical, it is likely that we 
can see these lines closer to the nucleus than optical narrow lines such as [O {\sc iii}] \citep{hss5}.

At least three of the mid-IR luminous NLRGs in our sample are known to have highly polarized 
broad emission lines. The NLRG 3C 234 has quasar-like mid-IR luminosity and a highly polarized broad 
H$\alpha$ line \citep{ant84}. The NLRG 3C 265 has quasar luminosity, highly polarized UV flux, and a 
highly polarized broad Mg {\sc ii} line \citep{tco98}. The low-redshift NLRG 3C 33 has
mid-IR luminosity comparable to the BLRG 3C 219, and highly polarized broad H$\alpha$ and
H$\beta$ \citep{cot99}. We have an ongoing program to obtain optical spectropolarimetry
of the rest of the sources in our sample. While the radio galaxies with highly polarized broad lines
are known to contain hidden type-1 AGNs, there were previously no reliable measurements
of the hidden AGN luminosities. Our mid-IR flux measurements yield rough calorimetric estimates 
of the hidden AGN luminosities, subject to uncertainties in SED, torus covering fraction, and
any mid-IR anisotropy.

\subsection{Mid-IR Weak Radio Galaxies}

The majority of radio galaxies in our sample (17/31 or $55\pm 13\%$) are relatively weak mid-IR 
sources. It is possible that some of the AGNs are highly obscured even at 15 $\mu$m because they are 
viewed through a very large dust column. A mid-IR source obscured by a nearly Compton-thick 
($\tau_\mathrm{e}=0.7$) disk with Galactic dust/gas ratio would have an equatorial extinction of 
$\sim 5$ mag (factor of 100) at 15 $\mu$m. Even in this case, any mid-IR emission above the disk might not 
be obscured. For example, there may be a contribution from dust in the narrow-line region (NLR), above the 
hole in the torus \citep[e.g. NGC 1068,][]{gpa05,bnm00}, tending to make the mid-IR emission more isotropic. 

For galaxies with redshift $z \le 0.22$, {\it Spitzer} can measure the flux at $\lambda = 30$ $\mu$m (rest), 
which should be more isotropic and less subject to extinction than the 15 $\mu$m flux. 
Nevertheless, 9/11 mid-IR weak galaxies in this redshift range are also weak at 30 $\mu$m, with 
$\nu L_\nu(30$ $\mu\mathrm{m}) <8 \times 10^{43}$ erg s$^{-1}$. The two exceptions are 
3C 61.1 and 3C 123, with  $\nu L_\nu(30$ $\mu\mathrm{m})=1.27 \pm 0.08$ and $1.4\pm 0.2\times 10^{44}$ erg s$^{-1}$,
respectively. In comparison 3C 452, the weakest mid-IR luminous NLRG in this redshift range, has 
$\nu L_\nu(30$ $\mu\mathrm{m})=8.63 \pm 0.09\times 10^{43}$ erg s$^{-1}$. Therefore, reclassifying the NLRGs by 
their luminosity at 30 $\mu$m would only gain us an additional 2 mid-IR luminous sources. This leads us to believe 
that most of the mid-IR weak radio galaxies truly lack a powerful accretion disk. Relatively low accretion 
power suggests, but does not prove, that some FR II jets may be driven by radiatively inefficient 
accretion flows or black hole spin-energy \citep{bbr84, m99}.

As we mentioned above, roughly half (9/17) of the mid-IR weak NLRGs are are classified as LEGs with weak optical 
[O {\sc iii}] emission. The [O {\sc iii}] emission in these sources may be weak because there is no 
strong source of UV photons to power the NLR. Qualitatively similar conclusions have been drawn by other 
investigators \citep[e.g.,][]{hl79, ccc00,grw04}. Alternatively, the NLR may be partly or completely obscured in 
LEGs \citep{hss5}. It will be important to make a more quantitative assessment of the optical and IR emission line 
strengths, to evaluate the extinction and determine what UV luminosity is necessary to power the NLR in mid-IR weak
NLRGs.

\subsection{Radio Properties and Unification}

One of the major motivations for the radio galaxy and quasar unification theory is the deficit of lobe-dominant
vs. core-dominant FR II quasars \citep{b89}. Relativistic beaming models predict that there should be relatively 
more sources where the radio jet is beamed away and the high frequency radio core is weak. Core fluxes at 5 GHz
are tabulated for the 3CRR catalog by 
\cite{lrl03}\footnote{The online 3CRR catalog is available at http://www.3crr.dyndns.org.}. We identify the 
mid-IR luminous NLRGs in our sample with the missing lobe-dominant quasar population (Fig. 8). Their median core 
to lobe ratio is $R_\mathrm{c}=\nu L_\nu($core, 5 GHz$)/ \nu L_\nu(178$ MHz$)=5$, while the 
median $R_\mathrm{c}=180$ for the quasars. Conversely, all of the mid-IR luminous sources with 
$R_\mathrm{c}>100$ are classified as quasars. We will perform a more detailed statistical analysis when the 
Spitzer observations of our full sample are complete.

For most mid-IR weak NLRGs, we reject the possibility that the AGN and radio core are viewed in an 'off' 
state while the radio lobes are still active. A 5 GHz core is detected in 13/17 of the mid-IR weak radio galaxies 
(Fig. 8) and 13/14 of the mid-IR luminous galaxies. Furthermore, mid-IR weak and mid-IR luminous NLRGs have 
comparable core to total luminosity ratios of $R_\mathrm{c} = 1-100$. Deeper 5 GHz observations of the 5 
non-detected cores (in 3C 28, 153, 172, 315, and 319) will be necessary to determine whether or not they are in 
an off state (e.g., $R_\mathrm{c}<0.1$).

We find that FR II radio morphology is not a reliable predictor of nuclear mid-IR luminosity for 
radio galaxies. {\it Contrary to the simple unification paradigm, not all narrow-line 
FR II galaxies host nuclei as powerful as quasars with matched radio lobe luminosity}. 
Unification theories must be modified to account for an additional population of mid-IR weak radio galaxies. 

Both intrinsic jet power and interaction with the interstellar and intergalactic medium are likely to 
be important for determining radio morphology. The break luminosity between FR I and FR II radio 
sources is found to increase with host galaxy optical luminosity \citep{ol94,b96}. The existence of 
radio sources with hybrid FR I/II morphology also points to the importance of environmental effects in 
determining radio morphology \citep{gkw00,gmk05}. Furthermore, most FR I radio jets are one-sided, 
relativistic, and narrowly collimated on sub-parsec scales, just like FR IIs, and decollimate
only on kpc scales \citep[e.g. M87,][]{jbl99}. 

Contrary to a common misconception, not all FR I sources have radiatively inefficient nuclei. For 
example, Centaurus A is persuasively argued to have a powerful hidden AGN \citep{wa04}, and the BLRG 
3C 120 is a well-known FR I source.  Deep VLA observations of the optically luminous, 'radio-quiet' 
quasar E1821+643 demonstrate that it has an FR I radio morphology \citep{br01}. These and other cases 
\citep{ant01} demonstrate that many powerful AGNs are FR Is. 

The observed variation in radio morphology and a wide range in AGN radio-loudness \citep{kss89} do not 
necessarily require a weak coupling between jet power and accretion power, but may demonstrate that 
multiple factors are at work. At least five parameters may be necessary to theoretically unify all AGN 
types: black hole mass, black hole spin, accretion rate, radiative efficiency, and viewing angle. There
is still much work ahead before we completely understand how basic physical parameters regulate the 
activity of supermassive black hole systems. Models that tie jet production to accretion onto a 
spinning black hole are particularly promising \citep[e.g.,][]{m99,hk06}.

Much progress has been made in understanding the aspect-dependent 
appearance of AGN disks and jets, as a consequence of relativistic beaming and obscuration \citep{up95}.
Our {\it Spitzer} observations confirm that many FR II radio galaxies would appear as powerful quasars if 
viewed from an unobscured direction (e.g. along the radio axis). However, just as many FR II radio galaxies 
would not. Our {\it Spitzer} observations also demonstrate that powerful radio jets may be produced even by 
mid-IR weak AGN. A powerful, luminous accretion disk is not always necessary to produce a highly collimated, 
relativistic jet.

\section{Conclusions}

(1.) We report on a large {\it Spitzer} spectroscopic survey of 3CRR FR II radio sources with $S_{178}>16.4$ Jy 
at $z<1$. We find strong mid-IR emission from $45 \pm 12\%$ (14/31) of NLRGs, which have luminosities comparable 
to matched BLRGs and quasars. Other indicators including high-ionization mid-IR lines and highly polarized broad 
emission lines confirm that some of these sources contain hidden quasar or BLRG nuclei. This demonstrates the power
of {\it Spitzer} IRS for unveiling hidden quasars and estimating their luminosities. 

(2.) We present {\it Spitzer} spectra of the 14 mid-IR luminous radio galaxies. In most cases, the mid-IR 
continuum bump from 3-30 $\mu$m can be produced by a distribution of hot dust with temperatures in the range 
210-$660$ K. These high temperatures are most likely maintained by hidden AGNs. The silicate 
absorption trough at 9.7 $\mu$m has an apparent optical depth of $\tau=0-0.2$ in most cases, 
consistent with dust temperatures decreasing outward from the center of a dusty torus. Two
sources, 3C 55 and 433, have deeper silicate troughs which may be produced by additional cool dust 
in the host galaxy.

(3.) However, not all FR II radio galaxies emit strongly in the mid-IR. Contrary to single-population
unification schemes, the majority of narrow-line radio galaxies in our sample (17/31 or $55 \pm 13\%$)
have weak or undetected mid-IR emission compared to matched quasars and BLRGs, with  
$\nu L_\nu(15$ $\mu\mathrm{m}) < 8 \times 10^{43}$ erg s$^{-1}$.
For a few sources, this may possibly be the result of anisotropic torus emission viewed 
through a large column density of dust. However, it is likely that most of the weakest sources do not
contain a powerful accretion disk. These may be truly nonthermal, jet-dominated AGNs, where the jet is powered by 
a radiatively inefficient accretion flow or black hole spin-energy rather than energy extracted from an 
accretion disk.

\acknowledgements

This work is based on observations made with the {\it Spitzer} Space Telescope, which is operated 
by the Jet Propulsion Laboratory, California Institute of Technology under NASA contract 1407. 
We have also made use of the NASA/IPAC Extragalactic Database (NED) which is operated by the Jet Propulsion 
Laboratory, California Institute of Technology, under contract with NASA. Support for this research was 
provided by NASA through an award issued by JPL/Caltech. We thank Dave Meier, Lee Armus, Bill Reach, and the
anonymous referee for their helpful input and comments on the manuscript.

\begin{deluxetable}{cccccccc}
\tablecaption{Mid-IR Luminous Sources}
\tablewidth{0pt}
\tablehead{
\colhead{3C} & \colhead{type\tablenotemark{a}}
             & \colhead{z} 
             & \colhead{$F_7$\tablenotemark{b}} 
             & \colhead{$F_{15}$\tablenotemark{c}} 
             & \colhead{log $\nu L_{15}$ \tablenotemark{d }}
             & \colhead{$\alpha$ \tablenotemark{e}} 
             &\colhead{S178\tablenotemark{f}}
}
 
\startdata
175   & QSR  & 0.7700 & 6.96 $\pm$ 0.07 & 21.6 $\pm$ 0.7 & 45.83 &1.49 $\pm$ 0.04 & 19.2 \\ 
196   & QSR  & 0.871  & 8.0  $\pm$ 0.1  & 22.9 $\pm$ 0.6 & 45.96 &1.38 $\pm$ 0.04 & 74.3 \\ 
216   & QSR  & 0.6703 & 9.8  $\pm$ 0.1  & 28.7 $\pm$ 0.6 & 45.83 &1.41 $\pm$ 0.03 & 22.0 \\ 
219   & BLRG & 0.1744 & 3.6  $\pm$ 0.1  & 11.2 $\pm$ 0.4 & 44.21 &1.50 $\pm$ 0.06 & 44.9 \\ 
254   & QSR  & 0.7361 & 6.02 $\pm$ 0.08 & \nodata        & 45.56:&\nodata         & 21.7 \\ 
263   & QSR  & 0.646  & 13.6 $\pm$ 0.1  & 29.8 $\pm$ 0.8 & 45.81 &1.03 $\pm$ 0.04 & 16.6 \\ 
275.1 & QSR  & 0.5551 & 3.04 $\pm$ 0.07 & \nodata        & 44.76:&\nodata         & 19.9 \\ 
325   & QSR  & 0.8600 & 1.17 $\pm$ 0.07 &  4.1 $\pm$ 0.2 & 45.20 &1.6  $\pm$ 0.1  & 17.0 \\ 
380   & QSR  & 0.6920 & 15.1 $\pm$ 0.1  & 40.4 $\pm$ 1.2 & 46.00 &1.29 $\pm$ 0.04 & 64.7 \\ 
382   & BLRG & 0.0579 & 86.1 $\pm$ 0.3  &114.  $\pm$ 2.  & 44.24 &0.37 $\pm$ 0.02 & 21.7 \\
390.3 & BLRG & 0.0561 & 56.8 $\pm$ 0.5  &164.  $\pm$ 4.  & 44.37 &1.39 $\pm$ 0.03 & 51.8 \\ 
\hline
33    & HEG  & 0.0597 & 19.8 $\pm$ 0.3  & 75.  $\pm$ 2.  & 44.08 &1.75 $\pm$ 0.04 & 59.3 \\ 
55    & HEG  & 0.7348 & 4.7  $\pm$ 0.1  & 23.3 $\pm$ 0.7 & 45.82 &2.11 $\pm$ 0.06 & 23.4 \\
172   & HEG  & 0.5191 & 0.40 $\pm$ 0.04 &  1.5 $\pm$ 0.2 & 44.31 &1.7  $\pm$ 0.2  & 16.5 \\
220.1 & HEG  & 0.610  & 0.77 $\pm$ 0.05 &  2.4 $\pm$ 0.1 & 44.67 &1.5  $\pm$ 0.1  & 17.2 \\
234   & HEG\tablenotemark{g} & 0.1848 & 86.  $\pm$ 2.   & 239. $\pm$ 3.  & 45.59 &1.35 $\pm$ 0.03 & 34.2 \\ 
244.1 & HEG  & 0.4280 & 3.50 $\pm$ 0.09 & 14.4 $\pm$ 0.3 & 45.13 &1.86 $\pm$ 0.04 & 22.1 \\
263.1 & HEG  & 0.8240 & 0.63 $\pm$ 0.07 & 2.7  $\pm$ 0.1 & 44.98 &1.9  $\pm$ 0.2  & 19.8 \\
265   & HEG  & 0.8110 & 9.0  $\pm$ 0.1  & 21.1 $\pm$ 0.5 & 45.86 &1.12 $\pm$ 0.04 & 21.2 \\
268.1 & HEG  & 0.970  & $<0.5$          &  3.0 $\pm$ 0.2 & 45.17 &2.1  $\pm$ 0.2\tablenotemark{h} & 23.3 \\
280   & HEG  & 0.996  & 4.58 $\pm$ 0.09 & 13.2 $\pm$ 0.4 & 45.83 &1.39 $\pm$ 0.05 & 25.8 \\
330   & HEG  & 0.550  & 1.72 $\pm$ 0.06 &  6.4 $\pm$ 0.2 & 45.00 &1.72 $\pm$ 0.06 & 30.3 \\
381   & HEG  & 0.1605 & 11.1 $\pm$ 0.2  & 36.4 $\pm$ 0.8 & 44.65 &1.56 $\pm$ 0.05 & 18.1 \\
433   & HEG  & 0.1016 & 40.0 $\pm$ 0.4  & 98.  $\pm$ 1.  & 44.67 &1.18 $\pm$ 0.02 & 61.3 \\
452   & HEG  & 0.0811 & 11.3 $\pm$ 0.2  & 45.  $\pm$ 1.  & 44.13 &1.80 $\pm$ 0.04 & 59.3 \\
\enddata
\tablenotetext{a}{Optical spectral type \citep{lrh99,jr97}.}
\tablenotetext{b,c}{Flux densities (mJy) and $3\sigma$ upper limits at 7 $\mu$m (rest) and 
                    15 $\mu$m (rest) from {\it Spitzer} IRS.}
\tablenotetext{d}{~Logarithm of luminosity (erg s$^{-1}$) at 15 $\mu$m (rest). Values for 3C 254
                  and 3C 275.1 are extrapolated from  7 $\mu$m because the LL data are unavailable.}
\tablenotetext{e}{Spectral power law index for $F_\nu\sim \nu^{-\alpha}$ from 
                  7-15 $\mu$m (rest).}
\tablenotetext{f}{Radio flux density (Jy) at 178 MHz (observed) \citep{lrl83}, multiplied by
                  a factor of 1.09 to convert to the \cite{b77} standard flux scale.}
\tablenotetext{g}{The broad H$\alpha$ line visible in total flux is entirely scattered \citep{ant84}.} 
\tablenotetext{h}{Spectral power law index for 3C 268.1 measured from 8-15 $\mu$m (rest).}
\end{deluxetable}

\begin{deluxetable}{cccccccc}
\tablecaption{Mid-IR Weak Sources}
\tablewidth{0pt}
\tablehead{
\colhead{3C} & \colhead{type\tablenotemark{a}}
             & \colhead{z} 
             & \colhead{$F_7$\tablenotemark{b}} 
             & \colhead{$F_{15}$\tablenotemark{c}} 
             & \colhead{log $\nu L_{15}$ \tablenotemark{d }}
             & \colhead{$\alpha$ \tablenotemark{e}} 
             &\colhead{S178\tablenotemark{f}}
}
 
\startdata 
61.1  & HEG  & 0.1878 & 0.64 $\pm$ 0.06 &  3.0 $\pm$ 0.2 &   43.70 &2.0  $\pm$ 0.2  & 34.0 \\
192   & HEG  & 0.0597 & 1.28 $\pm$ 0.07 &  3.2 $\pm$ 0.2 &   42.71 &1.2  $\pm$ 0.1  & 23.0 \\
274.1 & HEG  & 0.4220 & 0.20 $\pm$ 0.05 &  $<$0.9        &$<$43.91 &\nodata         & 18.0 \\
300   & HEG  & 0.270  & 0.26 $\pm$ 0.04 &  0.7 $\pm$ 0.2 &   43.40 &1.3  $\pm$ 0.4  & 19.5 \\
315   & HEG  & 0.1083 & 0.97 $\pm$ 0.08 &  1.9 $\pm$ 0.2 &   43.01 &0.9  $\pm$ 0.2  & 19.4 \\
388   & HEG  & 0.0917 & 0.98 $\pm$ 0.06 & 0.84 $\pm$ 0.09&   42.66 &0.4  $\pm$ 0.2  & 26.8 \\
436   & HEG  & 0.2145 & 0.65 $\pm$ 0.05 & 1.5  $\pm$ 0.2 &   43.52 &1.1  $\pm$ 0.2  & 19.4 \\
438   & HEG  & 0.290  & 0.16 $\pm$ 0.04 &$<$0.45         &$<$43.27 &\nodata         & 48.7 \\
\hline
28    & LEG  & 0.1953 & 0.45 $\pm$ 0.06 &  $<$0.30       &$<$42.74 &\nodata         & 17.8 \\
123   & LEG  & 0.2177 & 1.07 $\pm$ 0.08 &  2.8 $\pm$ 0.4 &   43.81 &0.7  $\pm$ 0.2  &206.0 \\
153   & LEG  & 0.2769 & 0.29 $\pm$ 0.05 &  1.0 $\pm$ 0.2 &   43.59 &1.7  $\pm$ 0.3  & 16.7 \\
173.1 & LEG  & 0.2921 & 0.38 $\pm$ 0.04 &  0.6 $\pm$ 0.1 &   43.40 &0.6  $\pm$ 0.3  & 16.8 \\
288   & LEG  & 0.2460 & 0.40 $\pm$ 0.05 &$<$0.60         &$<$43.25 &\nodata         & 20.6 \\
310   & LEG  & 0.0538 & 0.81 $\pm$ 0.07 &  0.73$\pm$ 0.1 &   41.98 &-0.1 $\pm$ 0.2  & 60.1 \\
319   & LEG  & 0.1920 & $<$0.15         &$<$0.27         &$<$42.68 &\nodata         & 16.7 \\
326   & LEG  & 0.0895 & 0.67 $\pm$ 0.09 &$<$0.39         &$<$42.16 &\nodata         & 22.2 \\
401   & LEG  & 0.2011 & 0.25 $\pm$ 0.05 &  0.8$\pm$  0.2 &   43.17 &1.5 $\pm$ 0.4   & 22.8 \\
\enddata
\tablenotetext{a-f}{See Table 1.}

\end{deluxetable}

\begin{deluxetable}{ccc}
\tablecaption{Silicate Trough}
\tablewidth{0pt}
\tablehead{
\colhead{3C} & \colhead{EW$_{9.7}$ \tablenotemark{a}} & \colhead{$\tau_{9.7}$ \tablenotemark{b}} 
}
 
\startdata
33    & -0.222 $\pm$ 0.007 &  0.14  $\pm$ 0.02  \\   
55    & -1.51  $\pm$ 0.03  &  0.9   $\pm$ 0.1   \\
172   & $>-0.4$            &  \nodata           \\
220.1 & $>-0.3$            &  \nodata           \\
234   & -0.028 $\pm$ 0.006 &  0.019 $\pm$ 0.008 \\
244.1 & -0.23  $\pm$ 0.03  &  0.18  $\pm$ 0.08  \\
263.1 & -1.3   $\pm$ 0.2   &  \nodata           \\
265   & -0.11  $\pm$ 0.03  &  0.04  $\pm$ 0.06  \\
268.1 & -0.4   $\pm$ 0.1   &  0.2   $\pm$ 0.5:  \\
280   & -0.23  $\pm$ 0.02  &  0.16  $\pm$ 0.09  \\
330   & -0.41  $\pm$ 0.06  &  0.1   $\pm$ 0.1   \\
381   & -0.19  $\pm$ 0.02  &  0.07  $\pm$ 0.02  \\
433   & -1.30  $\pm$ 0.01  &  0.71  $\pm$ 0.07  \\
452   & -0.196 $\pm$ 0.009 &  0.10  $\pm$ 0.02  \\
\enddata
\tablenotetext{a}{The 9.7 $\mu$m silicate trough  (rest) equivalent width in $\mu$m.}
\tablenotetext{b}{Apparent 9.7 $\mu$m silicate optical depth, averaged over 9.2-10.2 $\mu$m (rest).}
\end{deluxetable}

\begin{deluxetable}{ccccccccc}
\tablecaption{Emission Lines\tablenotemark{a}}
\tablewidth{0pt}
\tablehead{
\colhead{3C} & \colhead{[Ne {\sc vi}]} & \colhead{[S {\sc iv}]} & \colhead{[Ne {\sc ii}]}
             & \colhead{[Ne {\sc v}]}  & \colhead{[Ne {\sc iii}]} & \colhead{[S {\sc iii}]}
             & \colhead{[Ne {\sc v}]}  & \colhead{[O {\sc iv}]} \\
\colhead{ }  & \colhead{$\lambda$ 7.65}  & \colhead{10.51} & \colhead{12.81} 
             & \colhead{14.3} & \colhead{15.55}  & \colhead{18.71} 
             & \colhead{24.31} & \colhead{25.89}  
}
\startdata
33    & 2.8(0.2)& 1.2(0.1)& 3.9(0.2)& 2.0(0.3)& 5.3(0.2)& 2.5(0.4)& 1.6(0.2)& 8.1(0.2)\\    
      & 0.022   & 0.012   & 0.037   & 0.019   & 0.051   & 0.028   & 0.029   & 0.159   \\  
55    &1.82(.06)& 2.2(0.3)& $<0.5$  & 1.1(0.2)& 2.0(0.3)& $<0.4$  & \nodata & \nodata \\
      & 0.062   & 0.17    & $<0.01$ & 0.036   & 0.068   & $<0.01$ & \nodata & \nodata \\
172   & $<0.1$  & $<0.2$  & $<0.3$  & $<0.6$  & $<0.6$  & $<0.8$  & \nodata & \nodata \\
      & $<0.08$ & $<0.1$  & $<0.08$ & $<0.2$  & $<0.2$  & $<0.3$  & \nodata & \nodata \\
220.1 & $<0.2$  & 0.5(0.2)& $<0.2$  & $<0.5$  &0.35(.09)& $<0.2$  & \nodata & \nodata \\
      & $<0.06$ & 0.21    & $<0.05$ & $<0.1$  & 0.11    & $<0.05$ & \nodata & \nodata \\ 
234   & 1.7(0.1)& 3.2(0.3)& 0.8(0.2)& 3.3(0.7)& 8.2(0.7)& $<3.$   & 3.9(0.7)& 7.5(1.0)\\ 
      & 0.0034  & 0.0077  & 0.0022  & 0.010   & 0.027   & $<0.01$ & 0.026   & 0.056   \\
244.1 & 0.7(0.2)& 0.9(0.2)& 1.4(0.2)& 0.6(0.4)& 0.3(0.2)&0.68(.06)& \nodata & \nodata \\
      & 0.033   & 0.041   & 0.067   & 0.033   & 0.02    & 0.043   & \nodata & \nodata \\
263.1 & $<0.6$  & $<0.8$  & 0.6(0.2)& $<0.6$  & 0.6(0.2)& \nodata & \nodata & \nodata \\
      & $<0.1$  & $<0.3$  & 0.19    & $<0.2$  & 0.23    & \nodata & \nodata & \nodata \\
265   & 0.6(0.2)& 1.6(0.4)& $<0.8$  & $<0.6$  & $<0.5$  & \nodata & \nodata & \nodata \\
      & 0.012   & 0.040   & $<0.02$ & $<0.02$ & $<0.02$ & \nodata & \nodata & \nodata \\
268.1 &0.49(.08)& $<0.4$  & $<0.4$  & $<0.8$  & $<0.2$  & \nodata & \nodata & \nodata \\
      & 0.15    & $<0.1$  & $<0.1$  & $<0.4$  & $<0.04$ & \nodata & \nodata & \nodata \\
280   & $<0.3$  & $<0.5$  & 0.3(0.1)& $<0.4$  & 0.9(0.2)& \nodata & \nodata & \nodata \\
      & $<0.01$ & $<0.02$ & 0.014   & $<0.02$ & 0.050   & \nodata & \nodata & \nodata \\
330   & 0.4(0.2)&0.64(.08)& $<0.5$  & 0.4(0.1)& 0.7(0.2)&0.19(.08)& \nodata & \nodata \\
      & 0.03    & 0.076   & $<0.05$ & 0.04    & 0.09    & 0.02    & \nodata & \nodata \\
381   & 1.3(0.1)&0.62(.07)& 0.6(0.2)& 0.7(0.3)& 1.2(.3) &1.49(.08)&0.56(.08)& 3.9(0.2)\\
      & 0.019   & 0.012   & 0.011   & 0.014   & 0.025   & 0.040   & 0.026   & 0.196   \\
433   & 3.1(0.3)& 2.2(0.2)& 1.9(0.3)& 2.7(0.3)& 5.2(0.3)& 1.2(0.8)& 2.3(0.5)& 7.9(0.4)\\
      & 0.014   & 0.023   & 0.012   & 0.020   & 0.042   & 0.013   & 0.028   & 0.105   \\
452   & $<0.5$  & 0.7(0.1)&2.11(.07)& $<0.3$  & 1.9(0.2)& 1.4(0.4)& $<0.3$  & 1.3(0.2)\\
      & $<.008$ & 0.012   & 0.034   & $<.005$ & 0.032   & 0.026   & $<0.01$ & 0.057   \\
\enddata
\tablenotetext{a}{Notes. For each source, emission line fluxes ($10^{-14}$ erg s$^{-1}$ cm$^{-2}$) or 
               2$\sigma$ upper limits are on the first line and rest equivalent widths ($\mu$m) 
               are on the second line. Emission line rest wavelengths ($\mu$m) are at the top of
               each column.}    
\end{deluxetable}

\begin{figure}[ht]
  \plotone{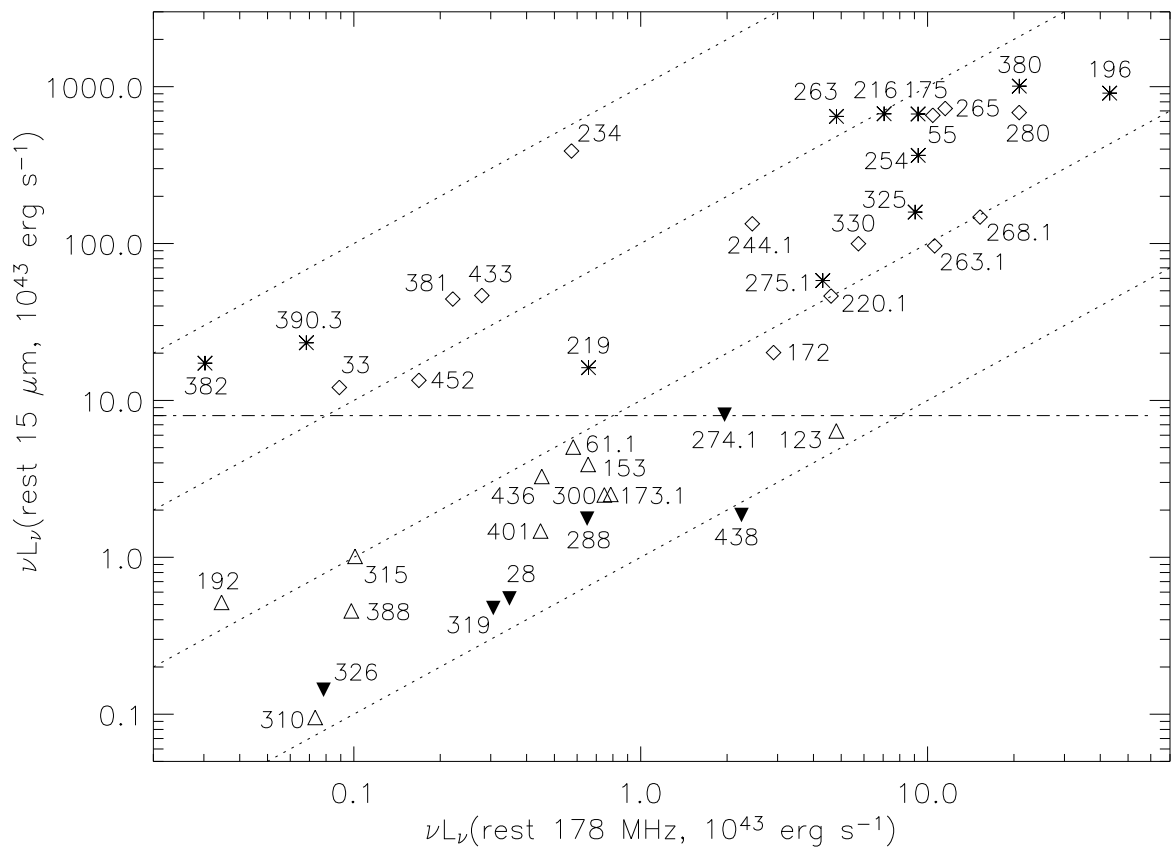}
  \figcaption{Comparison of 15 $\mu$m (rest) mid-IR power to 178 MHz (rest) radio power.
              The dotted lines indicate mid-IR to radio ratios of 1, 10, 100, and 1000.
              Quasars and BLRGs (asterisks) and mid-IR luminous NLRGs (diamonds) have mid-IR 
              luminosities $\nu L_\nu(15 \mu\mathrm{m}) > 8\times 10^{43}$ erg s$^{-1}$ (dot-dash 
              line). Mid-IR weak NLRGs (open triangles) have $\nu L_\nu(15 \mu\mathrm{m}) < 8\times 
              10^{43}$ erg s$^{-1}$. Filled triangles denote mid-IR weak NLRGs with  $3\sigma$ upper 
              detection limits at 15 $\mu$m.}
\end{figure} 

\begin{figure}[ht]
  \plotone{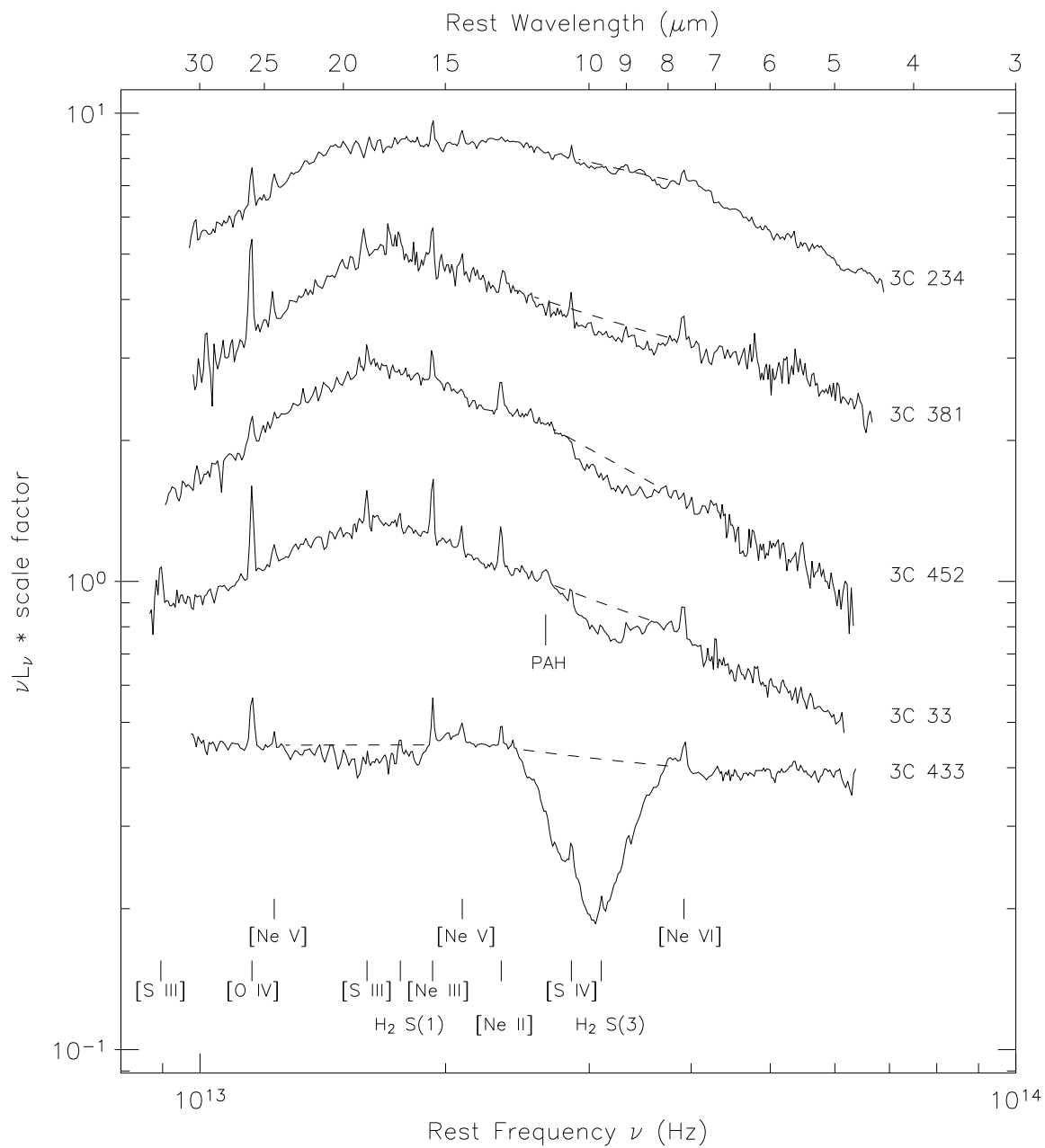}
  \figcaption{Spitzer spectra of  mid-IR luminous narrow-line radio galaxies ($z=0-0.2$), 
              ordered by 9.7 $\mu$m silicate trough optical depth. The continuum fits used to measure
              the silicate troughs are shown as dashed lines. The peak at 12-20 $\mu$m
              is characteristic of thermal dust emission. Several high-ionization forbidden
              emission lines, H$_2$ pure rotational lines, and the 11.3 $\mu$m PAH feature are labeled.} 
\end{figure}

\begin{figure}[ht]
  \plotone{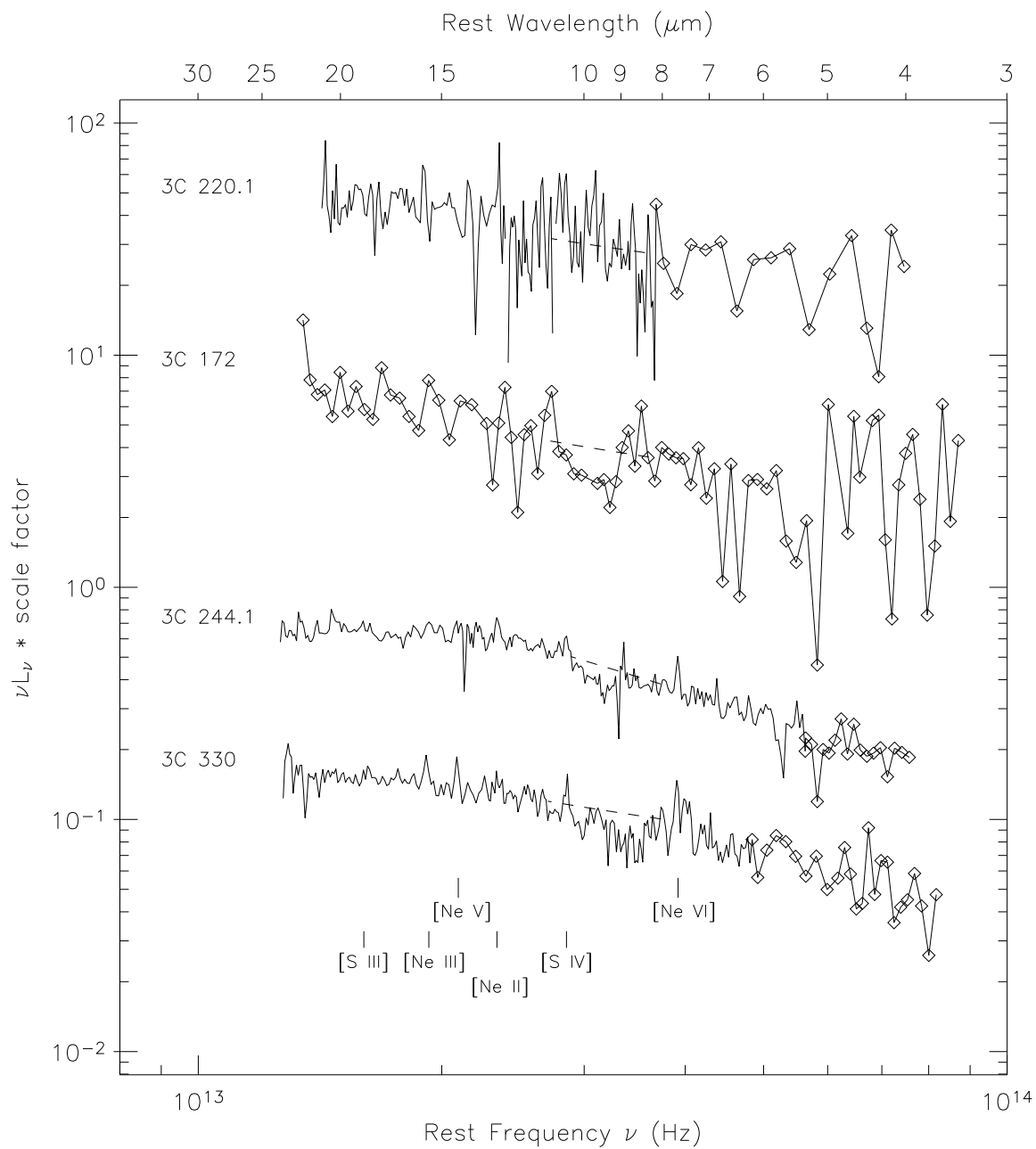}
  \figcaption{Spitzer spectra of mid-IR luminous NLRGs ($z=0.2-0.7$). 
              Some data are rebinned (diamonds) to improve S/N. Silicate absorption troughs (below dashed lines) 
              and forbidden emission lines are present in some sources.} 
\end{figure}

\begin{figure}[ht]
  \plotone{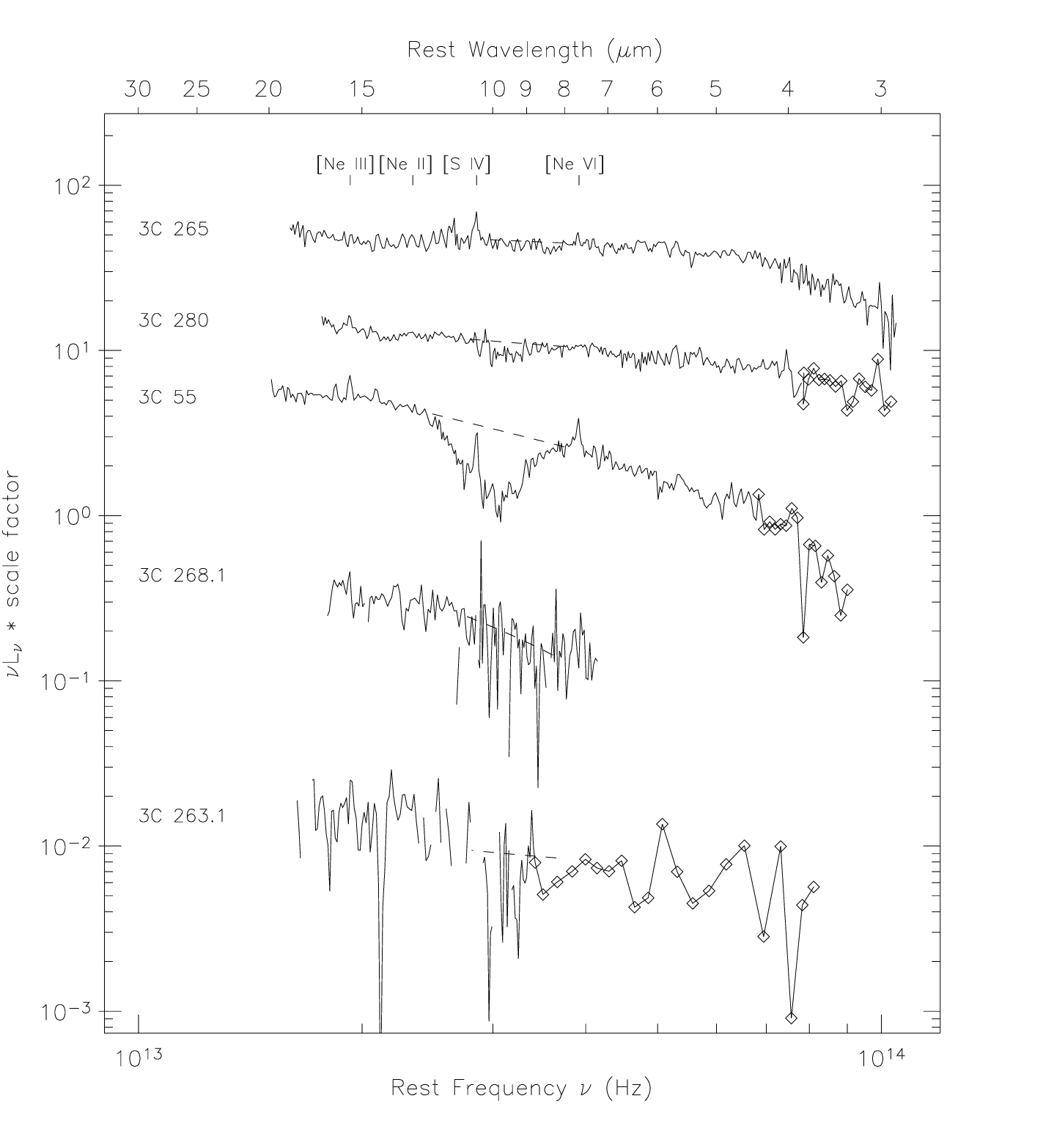}
  \figcaption{Spitzer spectra of mid-IR luminous NLRGs ($z=0.7-1.0$).
              Some data are rebinned (diamonds) to improve S/N. Some of the large downward spikes 
              and apparently missing data for 3C 263.1 and 268.1 are artifacts of plotting noisy 
              data on a logarithmic scale. Silicate absorption troughs (below dashed lines) 
              and forbidden emission lines are visible in some sources.} 
\end{figure}

\begin{figure}[ht]
  \plotone{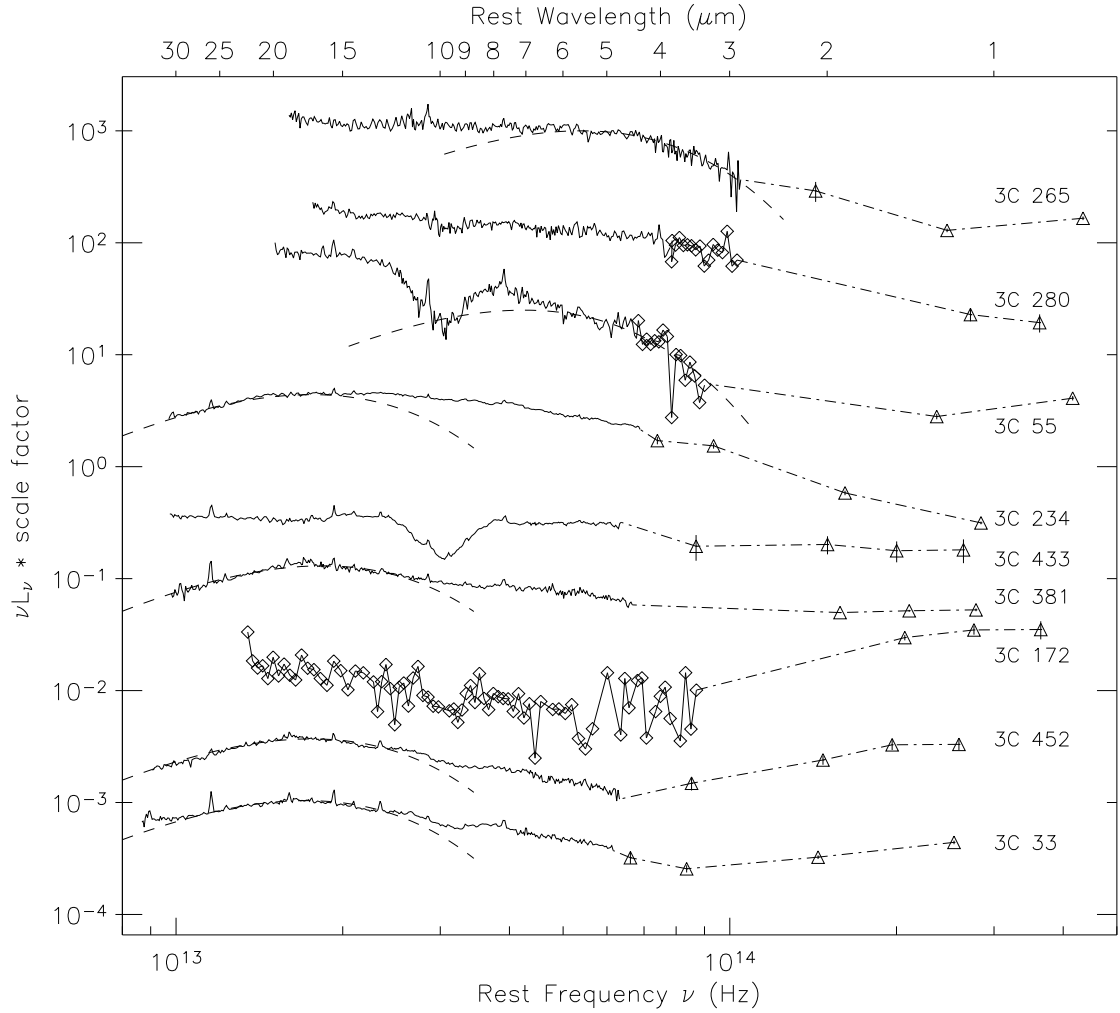}
  \figcaption{Spectral energy distributions of mid-IR-luminous NLRGs with near-IR photometry (triangles), 
              ordered by 15 $\mu$m luminosity. {\it Spitzer} data (solid curves) reveal excess mid-IR 
              emission relative to the near-IR stellar component. Blackbody fits (dashed 
              curves) to the low and high-frequency ends of the {\it Spitzer} spectra indicate dust 
              temperatures of 210-660 K.}
\end{figure}

\begin{figure}[ht]
  \plotone{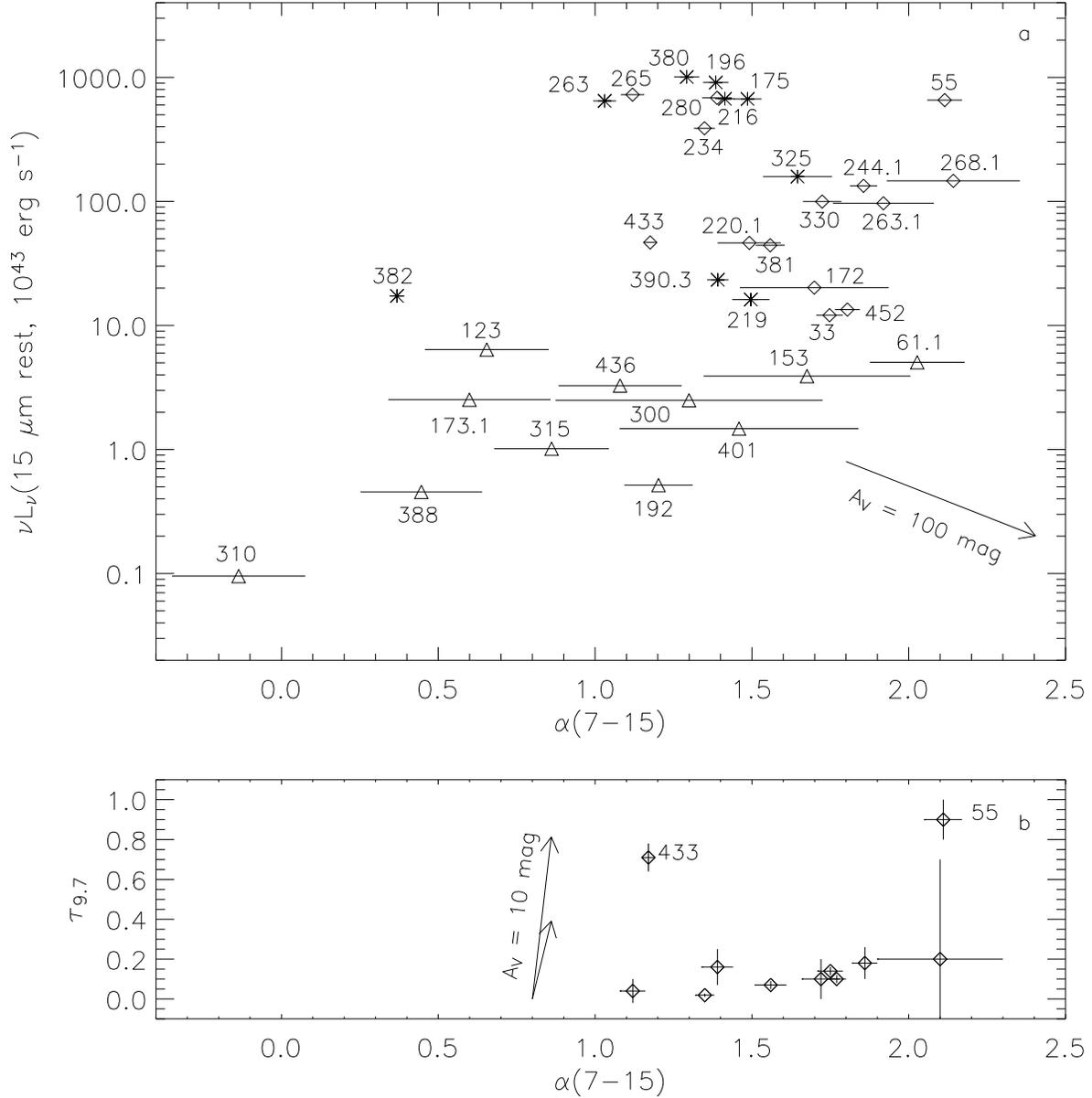}
  \figcaption{(a) Mid-IR luminosity vs. 7-15 $\mu$m spectral index for mid-IR luminous NLRGs (diamonds),
              mid-IR weak NLRGs (triangles), and quasars or BLRGs (asterisks). Mid-IR weak NLRGs with 
              $\alpha<1.0$ (e.g., 3C 310) have significant starlight from the host galaxy at 7 $\mu$m.
              The reddening vector corresponds to 100 magnitudes of optical (V band) extinction.
              (b) Silicate 9.7 $\mu$m apparent optical depth vs. 7-15 $\mu$m spectral index for mid-IR 
              luminous NLRGs with well-measured troughs. The two reddening vectors both correspond to 10 magnitudes 
              of optical (V band) extinction, for two different Galactic-type dust extinction curves (see text).}
\end{figure} 

\begin{figure}[ht]
  \plotone{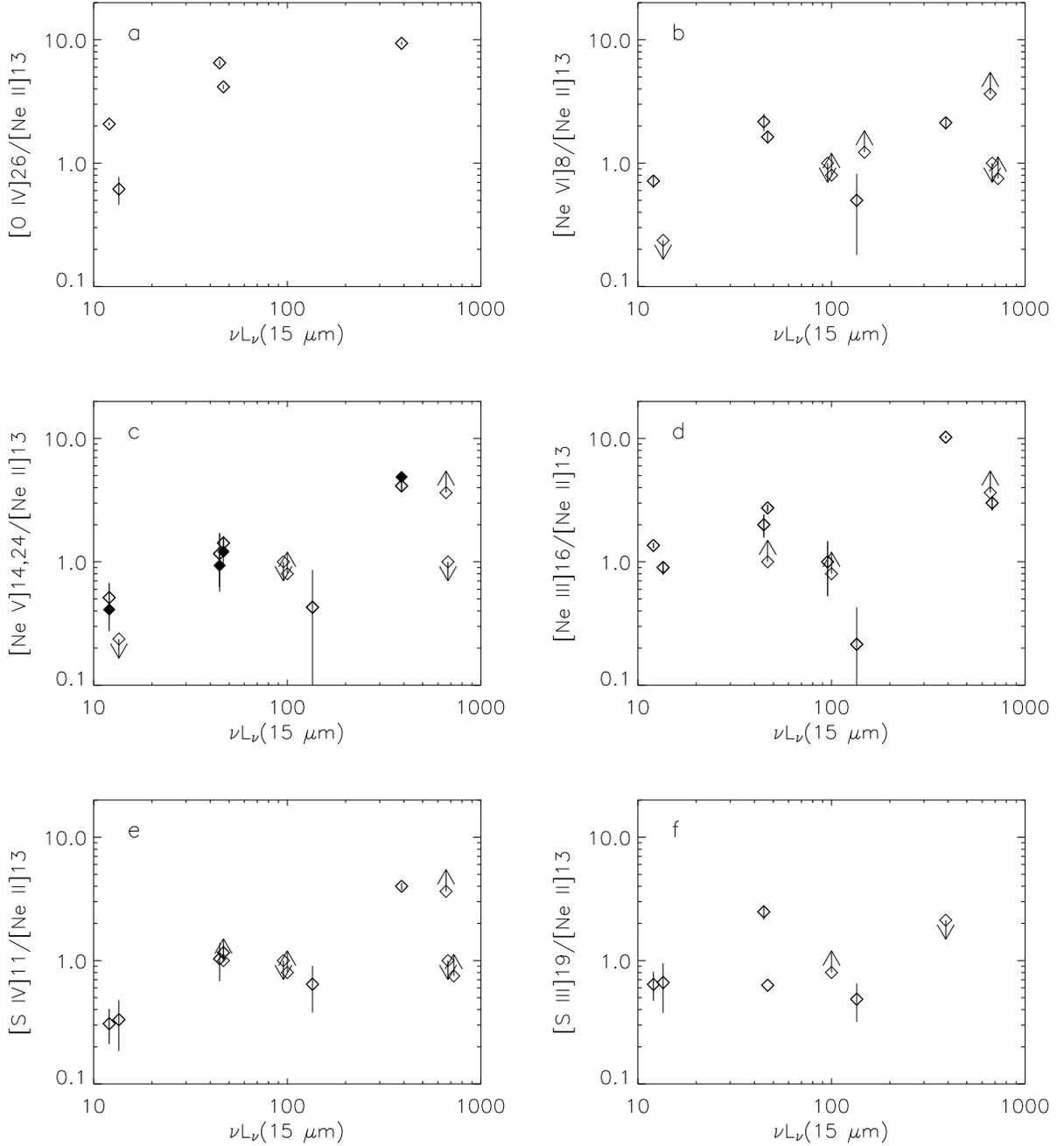}
  \figcaption{High-ionization emission line/ [Ne {\sc ii}] $\lambda$12.81 $\mu$m ratios vs. mid-IR luminosity 
              $\nu L_\nu($15 $\mu$m, rest) in units of $10^{43}$ erg s$^{-1}$, for mid-IR luminous NLRGs. 
              (a) [O {\sc iv}] $\lambda$25.89 $\mu$m,
              (b) [Ne {\sc vi}] $\lambda$7.65 $\mu$m,
              (c) [Ne {\sc v}] $\lambda$14.3 $\mu$m (open diamonds),
                  and [Ne {\sc v}] $\lambda$24.31 $\mu$m (filled diamonds),
              (d) [Ne {\sc iii}] $\lambda$15.55 $\mu$m,
              (e) [S {\sc iv}] $\lambda$10.51 $\mu$m, and (f) [S {\sc iii}] $\lambda$18.71 $\mu$m. 
              }
\end{figure} 

\begin{figure}[ht]
  \plotone{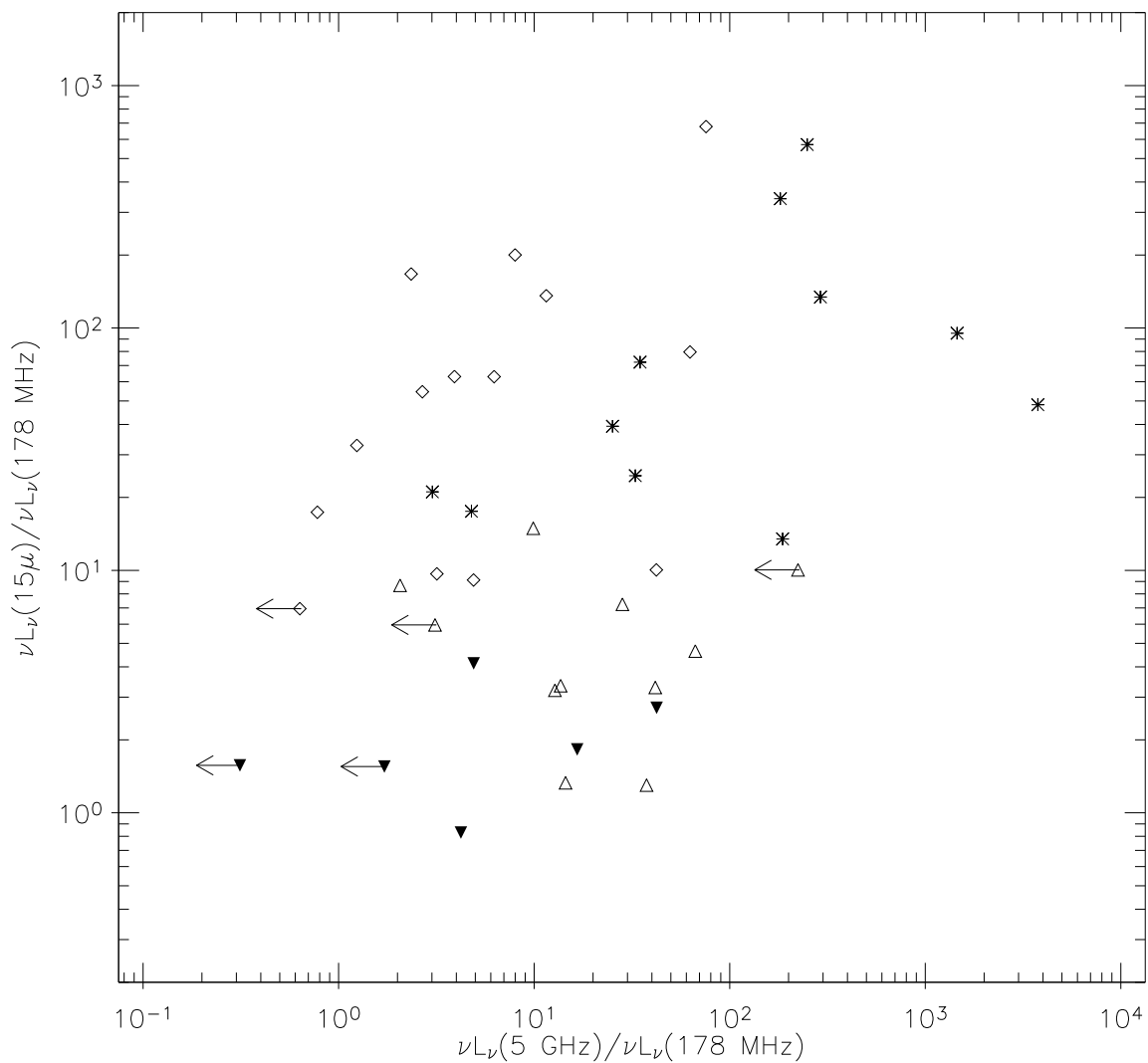}
  \figcaption{Mid-IR (15$ \mu$m, rest) luminosity vs. 5 GHz (observed) radio core luminosity, both normalized by 
              178 MHz (rest) radio lobe luminosity. The various symbols represent mid-IR weak NLRGs (triangles, 
              filled $=$ upper limit), mid-IR luminous NLRGs (diamonds), and quasars (asterisks). Upper limits at 
              5 GHz are indicated by arrows. The most core-dominant sources, 
              with $\nu L_\nu($5 GHz)/$\nu L_\nu($178 MHz)$>10^2$ are all quasars or BLRGs.}
\end{figure} 

\end{document}